\documentclass[aps,prl,twocolumn]{revtex4}%
\usepackage{makeidx}
\usepackage[english]{babel}
\usepackage[latin1]{inputenc}
\usepackage{latexsym}
\usepackage{amssymb}
\usepackage{amsmath}
\usepackage{graphicx}
\usepackage{graphics}
\usepackage{epsfig}
\usepackage{bbm}
\usepackage[dvips]{color}
\usepackage{amsfonts}%
\providecommand{\U}[1]{\protect\rule{.1in}{.1in}}
\begin{document}
\title{Wigner Crystallization in Rapidly Rotating 2D Dipolar Fermi Gases}
\author{M.A. Baranov}
\affiliation{Van der Waals-Zeeman Instituut, Unversiteit van Amsterdam, Valckenierstraat
65, 1018 XE Amsterdam, The Netherlands}
\affiliation{RRC Kurchatov Institute, Kurchatov sq. 1. 115582 Moscow, Russian Federation}
\author{H. Fehrmann}
\affiliation{Institut f\"ur Theoretische Physik, Leibniz Universit\"at Hannover,
Appelstra{ß}e 2, 30167 Hannover, Germany}
\author{M. Lewenstein}
\affiliation{ICREA and ICFO--Institut de Ciències Fot\`oniques, Castelldefels 08860, Spain}


\begin{abstract}
We study the competition between the Wigner crystal and the Laughlin liquid
states in an ultracold quasi two-dimensional rapidly rotating polarized
fermionic dipolar gas, and find that the Wigner crystal has a lower energy
below a critical filling factor. We examine the quantum crystal to liquid
transition for different confinements in the third direction. Our analysis of
the phonon spectra of the Wigner crystal with the account of phonon-phonon
interactions also shows the stability of the Wigner crystal for sufficiently
low filling factors ($\nu<1/7$).

\end{abstract}
\maketitle

Recently a remarkable progress has been made in experimental studies of
strongly correlated systems of ultra cold gases. A seminal breakthrough was
the observation of Mott insulator-superfluid transitions in Bose gases in
optical lattices \cite{Greiner}, followed by studies of low dimensional gases,
fermionic superfluidity or ultracold disordered gases (for recent reviews of
this very rapidly developing area see \cite{Bloch,ML}). A particularly
fascinating route toward creation of strongly correlated states is the one
that uses rapidly rotating gases. The rotation is formally equivalent to a
magnetic field (in the rotational frame), which reorganizes free-particle
states into discrete highly degenerate Landau levels. As a result, the
properties of the system become very sensitive to interparticle interactions,
in complete analogy to the Fractional Quantum Hall Effect (FQHE) for electrons
\cite{FQHF}, that exhibits a variety of strongly correlated states, among
which the Laughlin liquid \cite{Laughlin}, and the Wigner crystal
\cite{Wigner} are probably the most famous. This analogy has been pointed out
in the context of ultracold atomic gases with short range interactions in
Refs. \cite{Bose}, but an experimental observation of the FQHE in this case is
difficult, partially due to smallness of the energy gap separating a Laughlin
state from the so called quasi-hole excitations. It was recently argued
\cite{Baranov} that these problems may be overcome in rotating quasi-2D
polarized \textit{dipolar} gases, in which the interparticle interaction has a
long-range part that puts dipolar systems somewhat in between atomic systems
with short-range interactions and electron systems with Coulomb interactions.
As a result, the Laughlin state for the filling factor $\nu=1/3$ \footnote{The
filling factor $\nu$ denotes the fraction of occupied Landau levels. For a
rotating Bose gas it is the number of vortices per atom.} was predicted to be
incompressible with gapped excitations \cite{Baranov}.

Several other unexpected predictions have been already made for rotating
dipolar bosonic gases. The rotating dipolar condensates in the mean-field
regime exhibit novel forms of vortex lattices: square, "stripe crystal", and
"bubble crystal" lattices \cite{Cooper:2005,Komineas:2006}. On the other hand,
Rezayi \textit{et al.} \cite{Rezayi:2005} have recently shown that a small
amount of dipole-dipole interactions stabilizes the so called bosonic
Rezayi-Read strongly correlated state at $\nu=3/2$ whose excitations are both
fractional, and non-Abelian.

The dipolar interactions could also lead to a crystal ground state in a quasi
2D system similar to the electron Wigner crystal. In the case of an electron
gas, both in the presence and in the absence of a magnetic field, Wigner
crystal becomes energetically favorable and stable for low filling factors:
this is due to the fact that at low densities Coulomb interactions ($\sim1/R$,
where $R$ is a typical length scale) win over the kinetic energy ($\sim
1/R^{2}$) (cf. \cite{Girvin}). For non-rotating dipolar gases (where
interactions scale as $1/R^{3}$), it was argued \cite{Baranov} that the ground
state has a crystal order at \textit{high densities}, as shown in Refs.
\cite{Lozovik}. Remarkably, in the case of a rotating dipolar gas, the Wigner
crystal phase is expected at \textit{low densities} \cite{Baranov}. Therefore,
the dipolar gases, being to some extend similar to electrons, present
nevertheless a novel class of physical systems exhibiting quantum liquid to
crystal transition. This opens an unprecedented possibility for the
observation and detailed studies of this transition in clean and
well-controlled conditions. It is especially appealing in view of the recent
experimental realization of the dipolar Bose condensate of Chromium
\cite{Griesmaier:2005,PfauNature}, and progress in trapping and cooling of
dipolar molecules \cite{SpecialIssueEurphysD04}.

In this paper we investigate the existence of a Wigner crystal phase in a
rapidly rotating gas of polarized dipolar fermions by comparing it with the
competing liquid states. We then examine the stability of the Wigner crystal
by incorporating phonon-phonon interactions, and identify the border of
stability with the appearance of purely imaginary phonon frequencies. It
appears that the Wigner crystal is the stable ground state of the systems for
low filling factors, typically for $\nu<1/7$.

We consider a gas of fermionic polarized dipolar particles in a rotating
cylindrical trap, tightly confined and polarized along the axis of rotation.
For a sufficiently strong confinement in the axial direction, all particles
are in the ground state of the axial motion and we can write the many-body
wave function in the form
\[
\psi_{\mathrm{3D}}(\left\{  \mathbf{r}_{i},\xi_{i}\right\}  )=\psi
_{\mathrm{2D}}(\left\{  \mathbf{r}_{i}\right\}  )\left(  l\sqrt{\pi}\right)
^{-N/2}\exp(-\sum\nolimits_{i}\xi_{i}^{2}/2l^{2}),
\]
where $\mathbf{r}_{i}=\left(  x,y\right)  $ and $\xi_{i}$ are radial and axial
coordinates of the particles, respectively, $l$ is the extension in the axial
direction, $N$ is the total number of particles, and $\psi_{2D}$ is the wave
function of the system in the $xy$-plane. The confinement in the axial
direction also modifies the interparticle interaction in the $2D$-plane. For
two dipoles separated by a distance $r$ in the $2D$-plane, the effective
interaction potential is
\[
v_{\mathrm{2D}}(r)=\frac{d^{2}}{l^{3}\sqrt{2\pi}}\int_{0}^{\infty}d\xi
\sqrt{\frac{\xi}{(\xi+1)^{3}}}\exp(-\xi\frac{r^{2}}{2l^{2}})\quad.
\]
For $r\gg l$ the potential is $v_{\mathrm{2D}}\approx d^{2}/r^{3}$, while for
$r<l$ the effective potential increases logarithmically, $v_{\mathrm{2D}%
}\approx(d^{2}/l^{3})\sqrt{2/\pi}\ln(l/r)$. The wave function $\psi_{2D}$
obeys the Schrödinger equation with the effective 2D Hamiltonian that in the
rotational frame reads
\begin{equation}
H=-\sum_{i}\left(  \frac{\hbar^{2}}{2m}\Delta_{i}+\frac{m}{2}\omega_{\perp
}^{2}\mathbf{r}_{i}^{2}\right)  -\Omega L_{z}+V_{D}\quad,\label{H1}%
\end{equation}
where $\Omega$ is the rotational frequency, $\omega_{\perp}$ is the radial
trap frequency, $V_{D}=\sum_{i<j}v_{2D}\left(  |\mathbf{r}_{i}-\mathbf{r}%
_{j}|\right)  $ describes the dipole-dipole interparticle interaction, and
$L_{z}$ is the $z$-component of the total angular momentum. The Hamiltonian
(\ref{H1}) can be rewritten in the form
\begin{equation}
H=\sum_{i}\frac{1}{2m}\left(  -i\hbar\nabla_{i}-\mathbf{A}_{i}\right)
^{2}-\left(  \Omega-\omega_{\perp}\right)  L_{z}+V_{\mathrm{D}}%
\label{eq:Ham_mag}%
\end{equation}
with $\mathbf{A}_{i}=m\omega_{\perp}\mathbf{e}_{z}\times\mathbf{r}_{i}$ and,
therefore, it formally describes a system of charged particles in a constant
magnetic field with the cyclotron frequency $\omega_{c}=2\omega_{\perp}$. For
non-interacting particles ($V_{\mathrm{D}}=0$) in the regime of the critical
rotation ($\omega_{\perp}=\Omega$), the properties of the Hamiltonian
(\ref{eq:Ham_mag}) are well known; it's spectrum consists of highly degenerate
levels $E_{n}=\hbar\omega_{c}\left(  n+1/2\right)  $ called Landau levels. In
the following, we restrict ourself to a system of particles occupying only the
lowest Landau level.

The wave function of a non-correlated Wigner crystal on the lowest Landau
level is \cite{MakiZotos}:
\[
\Psi_{\mathrm{C}}(\{\mathrm{z}_{i}\}) \sim\mathcal{A}\prod_{i} \exp\left[
\frac{-1}{4l_{0}^{2}} ( \left\vert z_{i}-R_{i} \right\vert ^{2} +z_{i}%
R_{i}^{\ast}-z_{i}^{\ast}R_{i} ) \right]  ,
\]
where $z_{i}=x_{i}+iy_{i}$ is the complex representation of the $2D$ vector
$\mathbf{r}_{i}$, $R_{i}$ the (complex) lattice site, $l_{0}=\sqrt
{\hbar/m\omega_{c}}$ the magnetic length, and $\mathcal{A}$ the
antisymmetrization operator, which can be omitted for sufficiently low filling
factors $\nu=2\pi l_{0}^{2}n$. One can check that similar to the case of
classical dipoles, the energy is minimal for a triangular lattice with
particles centered at positions $\mathbf{R}_{i}=l_{1}\mathbf{b}_{1}%
+l_{2}\mathbf{b}_{2}$, where $l_{1,2}$ are integers, $\mathbf{b}_{1}=a\left(
0,1\right)  $, $\mathbf{b}_{2}=a\left(  \sqrt{3},1\right)  /2$, and $a$ is the
lattice constant determined by the density $n$ of the gas, $a^{2}=2/\sqrt
{3}n=4\pi l_{0}^{2}/\sqrt{3}\nu$. The energy of the crystal state equals
\begin{align*}
U_{\mathrm{C}}  &  = \left\langle \Psi_{\mathrm{C}} \vert V_{D} \vert
\Psi_{\mathrm{C} }\right\rangle \approx d^{2}n^{3/2}(0.2823 +0.2146\beta\\
&  + 0.3388\beta^{2}+0.7456\beta^{3}+2.0676\beta^{4}+\ldots),
\end{align*}
where $\beta=\pi n (2l_{0}^{2}-l^{2})$ and the first term corresponds to the
energy of classical point-like dipoles, $E_{\mathrm{cl}}=\sum\nolimits_{i<j}%
d^{2}/|l_{1}\mathbf{b}_{1}+l_{2}\mathbf{b}_{2}|^{3}=5.513d^{2}/a^{3}$.

The energy of the Laughlin state (for a small number of particles interacting
via dipolar forces, recent computations show a remarkable overlap of the exact
ground state with the Laughlin state \cite{Osterloh_exact}) characterized by
an odd integer $M=1/\nu$ is
\[
U_{\mathrm{L}}=\frac{\nu}{2}\int_{0}^{\infty}rg(r)v_{\mathrm{2D}}(r)dr,
\]
where $g(r)$ is the pair correlation function. We calculated this functions
and the energies for odd $M$ from $1$ to $19$ for a gas of $512$ particle
using a Monte Carlo method.

\begin{figure}[th]
\begin{center}
\includegraphics[width=8cm]{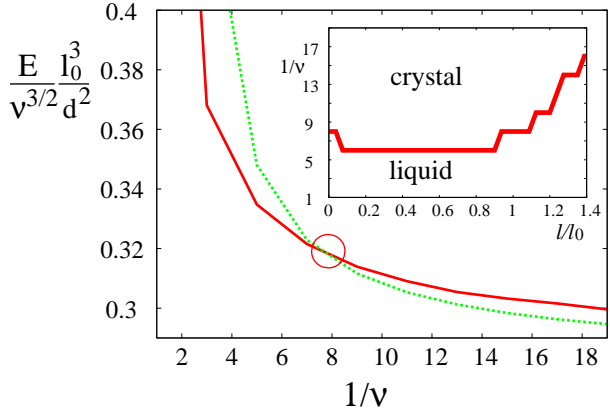}
\end{center}
\caption{Energy per particle for the Wigner crystal (dotted line) and for the
Laughlin state (solid line) as a function of the filling factor for $l=0$. The
insert shows the critical filling factor as a function of the extension in the
$z$-direction.}%
\label{Fig:Transition}%
\end{figure}

We now compare the energies of the Wigner crystal and of the Laughlin liquid
for different filling factors $\nu$ and extensions $l$ in the axial direction.
The results are presented in Fig. \ref{Fig:Transition}. They show that below
some critical value $\nu_{c}$, which depends on $l$, the Wigner crystal has a
lower energy than the Laughlin liquid and, therefore, for\ $\nu<\nu_{c}$ the
ground state is expected to be a crystal. Note that in obtaining the critical
filling factor we compare the energies of two simplest trail wave functions
for liquid and crystal states. A better estimate could follow from considering
the wave functions of quantum Hall liquids of composite fermions \cite{Jain}
with the filling factors closer to the critical one than primary Laughins
$1/M$, and of the correlated Wigner crystals \cite{Yi}. We, however, do not
expect a significant change of our result because for a dilute system the two
considered wave functions already catch the most important features of the
interparticle interaction. 

Let us now approach the liquid-crystal transition starting from the crystal
phase. For this purpose we consider the phonon spectrum with taking into
account phonon-phonon interactions (anharmonicity effects). The appearance of
purely imaginary phonon frequencies will indicate the instability of the
crystal and, therefore, the transition to a liquid state.

In the harmonic approximation the phonon eigenfrequencies can be obtained from
the dynamic equations for displacements $u_{\alpha\mathbf{l}}$ of particles
along the $\alpha$-axis from their equilibrium positions $\mathbf{R_{l}}$ in
the lattice \cite{Fukuyama75}
\begin{equation}
m\ddot{u}_{\alpha\mathbf{l}}=\sum\nolimits_{\beta\mathbf{l}^{\prime}}%
\Phi_{\alpha\mathbf{l},\beta\mathbf{l}^{\prime}}^{(2)}u_{\beta\mathbf{l}%
^{\prime}}+\omega_{c}\varepsilon_{\alpha\beta}\dot{u}_{\beta\mathbf{l}%
^{\prime}},\label{EQ:dm}%
\end{equation}
where $\Phi_{\alpha\mathbf{l},\beta\mathbf{l}^{\prime}}^{(2)}=\partial
^{2}U_{\mathrm{C}}/\partial R_{\alpha\mathbf{l}}\partial R_{\beta
\mathbf{l}^{\prime}}$ is the dynamical matrix, $\alpha,\beta=x,y$ and
$\varepsilon_{\alpha\beta}$ is the antisymmetric tensor, $\varepsilon_{xy}=1$.
The last term in Eq. (\ref{EQ:dm}) corresponds to the Coriolis (Lorentz) force
due to the rotation (magnetic field).

\begin{figure}[ptb]
\begin{center}
\includegraphics[width=8cm]{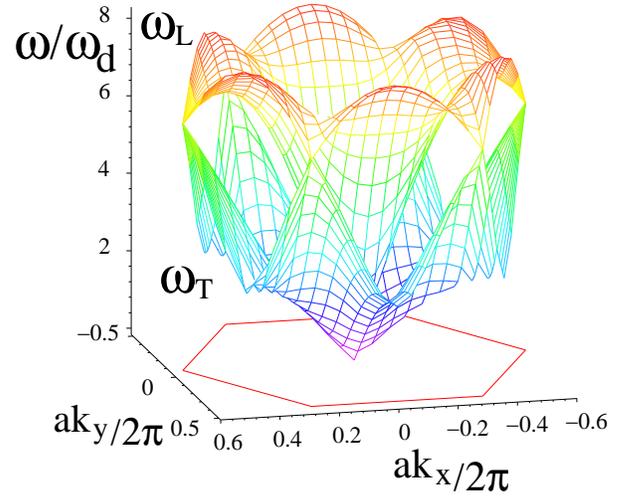}
\end{center}
\caption{Energy of the transverse $\omega_{T}$ (lower surface) and
longitudinal $\omega_{L}$ phonons.}%
\label{Fig:Disp}%
\end{figure}

Without rotation, $\omega_{c}=0$, Eq. \ref{EQ:dm} in the quasimomentum
representation reads:
\[
\omega^{2}\tilde{u}_{\alpha}(\mathbf{k})=\sum\nolimits_{\beta}F_{\alpha\beta
}(\mathbf{k})\tilde{u}_{\beta}(\mathbf{k}),
\]
where $mF_{\alpha\alpha^{\prime}}(\mathbf{k})=
\sum\nolimits_{\mathbf{l}}\exp\left[  -i\mathbf{k(R_{l}-R_{l^{\prime}}%
)}\right]  \Phi_{\alpha\mathbf{l},\beta\mathbf{l}^{\prime}}^{(2)}$ is the
Fourier transform of $\Phi_{\alpha\mathbf{l},\beta\mathbf{l}^{\prime}}^{(2)}$,
and the eigenvalues $\omega_{s}^{2}(\mathbf{k})$, $s=T,L$ of $F_{\alpha
\alpha^{\prime}}(\mathbf{k})$ determines the frequencies of the transversal
($T$) and longitudinal ($L$) phonons. The phonon frequencies for $l=0$ are
shown in Fig. \ref{Fig:Disp}. They are linear in $k$ for $k\ll a^{-1}$:
$\omega_{T}\approx(3/\sqrt{8})\omega_{d}ak$ and $\omega_{L}\approx\sqrt
{11}\omega_{T}$, where $\omega_{d}=\sqrt{d^{2}/ma^{5}}$ sets the typical value
for the frequency of phonons.

For a rotating crystal, $\omega_{c}\neq0$, the phonon spectrum becomes
\cite{Chaplik,Fukuyama75}
\[
\omega_{\pm}^{2}=\frac{\omega_{L}^{2}+\omega_{T}^{2}+\omega_{c}^{2}}{2}%
\pm\frac{1}{2}\sqrt{(\omega_{L}^{2}+\omega_{T}^{2}+\omega_{c}^{2})^{2}%
-4\omega_{L}^{2}\omega_{T}^{2}}.
\]
In the case $\omega_{c}\gg\omega_{s}$, one has $\omega_{+}\approx\omega_{c}$
and $\omega_{-}\approx\omega_{L}\omega_{T}/\omega_{c}$.

Higher orders (anharmonic) terms in the expansion of the energy of the crystal
with respect to the displacements of particles from their equilibrium
positions result in phonon-phonon interactions and, therefore, in the
renormalization of the phonon frequencies. At a given quasimomentum
$\mathbf{k}$, the renormalized frequencies correspond to the poles of the
Fourier transform $G_{\alpha\beta}(\omega,\mathbf{k})$ of the phonon Green
function $G_{\alpha\beta}(t,\mathbf{l}-\mathbf{l}^{\prime})=i\left[
\theta(t)\langle u_{\alpha\mathbf{l}}(t)u_{\beta\mathbf{l}^{\prime}}%
(0)\rangle+\theta(-t)\langle u_{\beta\mathbf{l}^{\prime}}(0)u_{\alpha
\mathbf{l}}(t)\rangle\right]  $ with $\theta(t)$ being the step function (for
more details on the usage of Green functions for phonons in crystals see e.g.
\cite{Boettger}). The Green function in the harmonic approximation is
$G_{\alpha\beta}^{(0)-1}(\omega,\mathbf{k})=\{\sum_{s=T,L}\mathcal{M}%
_{\alpha\beta}^{(s)}[\omega_{s}^{2}(\mathbf{k})-\omega^{2}]-i\varepsilon
_{\alpha\beta}\omega_{c}\omega\}m/\hbar$, where $\mathcal{M}_{\alpha\beta
}^{(s)}=e_{\alpha}^{(s)}e_{\beta}^{(s)}$ is the projector to the eigenmode $s$
with the polarization $\mathbf{e}^{(s)}$, and the poles are at $\omega
=\omega_{\pm}(\mathbf{k})$. In the presence of phonon-phonon interactions, the
Green function obeys the Dyson equation%

\[
G_{\alpha\beta}^{-1}(\omega,\mathbf{k})=G_{\alpha\beta}^{(0)-1}(\omega
,\mathbf{k})-\Sigma_{a\beta}(\omega,\mathbf{k}),
\]
where the phonon self-energy function $\Sigma$ incorporates all the effects of
phonon-phonon interactions.

\begin{figure}[ptb]
\includegraphics[width=6cm]{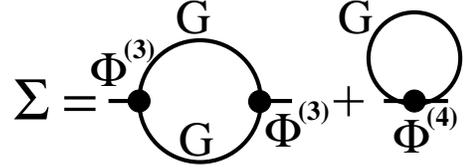}\caption{One-loop diagrams for the
phonon self-energy.}%
\label{Fig:Sigma}%
\end{figure}We calculate the phonon self-energy in the one-loop approximation
that has proven its reliability in the case of an electron Wigner crystal in a
strong magnetic field \cite{Badenov,Lozovik1,Lozovik2}. This corresponds to
taking into account only the fourth $\Phi^{(4)}\sim\partial^{4}U_{\mathrm{C}%
}/\partial R^{4}$ and the square of the third $\Phi^{(3)}\sim\partial
^{3}U_{\mathrm{C}}/\partial R^{3}$ order anharmonic terms, see Fig
\ref{Fig:Sigma}. We then solve the Dyson equation numerically by successive
iterations until we either obtain a self-consistent solution or the iteration
breaks down due to the appearance of purely imaginary phonon frequencies. The
latter indicates the instability of the lattice. The results of our
calculations can be summarized as follows. At very low filling factors the
phonon-phonon interactions do not play any significant role and the system
behaves harmonically. However, the effects of anharmonicity become
progressively important with increasing the filling factor, and, finally, at
some critical filling factor $\nu_{c}$ we find purely imaginary frequencies
and, hence, the crystal ceases to exist. We always observe this phonon
instability for $k\rightarrow0$. This indicates the breakdown of the
crystalline order and rules out any possible structural phase transition.
Therefore, the value $\nu_{c}$ sets the upper bound for the stability of the
crystal. The critical filling factor $\nu_{c}$ depends on the confinement in
axial direction as well as on the ratio of $\omega_{d}$ and $\omega_{c}$ that
measures the strength of the dipole-dipole interaction relative to the Landau
level spacing. This dependence for $l=0$ can be approximated as $\nu_{c}%
^{-1}=4.33\exp(-0.0021\omega_{c}/\omega_{d})+5.77$. Note that for $\omega
_{c}\gg\omega_{d}$ the critical filling factor $\nu_{c}$ becomes insensitive
to the strength of the interparticle interaction. In this limit the Wigner
crystal is stable for $\nu<\nu_{c}=0.174$, which is in a good agreement with
our previous energetic consideration for the tight confinement along the
$z$-axis, $l<l_{0}$.

We also calculate the Lindemann parameter $\gamma$ -- the ratio of the average
displacement of a particle in the lattice from its equilibrium position to the
lattice spacing $a$, $\gamma=\sqrt{\left\langle \mathbf{u}^{2}\right\rangle
}/a$. For $\omega_{c}\gg\omega_{d}$ we find $\gamma=0.28$ that is within the
range of values of the Lindemann parameter for various 2D crystals.

As we see, the crucial parameter that controls the ground state of the system
and, therefore, drives the quantum melting transition is the filling factor.
The way of manipulating the filling factor in experiments with rotating gases
depends on an experimental setup. In the case of a critical rotation with an
extra (quartic) confinement \cite{Dalibard} this can be achieved by changing
the number of particles (the size of the system is fixed by an extra
confinement). For a purely harmonic confinement and \textquotedblleft under
critical\textquotedblright\ rotations \cite{Abo-Shaeer2001a,Cornell}, the
filling factor can be changed by varying the difference $\Omega-\omega_{\perp
}$. In this setup, the size of the system and, therefore, the filling factor
is determined by a competition between the interparticle interaction
$V_{\mathrm{D}}$ and the "tilting" term $\left(  \Omega-\omega_{\perp}\right)
L_{z}$ (see Eq. \ref{eq:Ham_mag}). In both cases, the appearance of a crystal
order could be detected by studying the shot noise correlations using the
Hanbury Brown and Twiss effect \cite{HBT} in a similar way as it was used to
observe the Mott insulator-superfluid transition in a lattice Bose gas
\cite{BlochHBT}.

In conclusion, we show the existence of the Wigner crystal state for low
filling factors in a rapid rotating 2D dipolar Fermi gas. The critical filling
factor, at which the quantum melting of the crystal into a quantum Hall liquid
takes place, depends on the confinement in the axial direction and on the
strength of the dipole-dipole interaction, but typically one expects a liquid
ground state for $\nu>1/7$.

We wish to acknowledge N. Barberàn, N.R. Cooper, J.M.F. Gunn, E. Jeckelmann,
A.H. MacDonald, K. Osterloh, and R. Wimmer for helpful discussions. This work
was supported by the Nederlandse Stichting voor Fundamenteel Onderzoek der
Materie (FOM), the Russian Foundation for Fundamental Research, the Deutsche
Forschungsgemeinschaft (SPP1116, SFB 407, GK 282), ESF PESC "QUDEDIS", EU IP
"SCALA", Spanish MEC (FIS 2005-04627 and Consolider-Ingenio 2010 "QOIT").

\end{document}